\documentclass[11pt,a4paper,iop]{emulateapj}

\newcommand{\co}{$^{12}$CO}
\newcommand{\coline}{\co~J=3-2}

\usepackage{epsfig}
\usepackage{amsmath}
\usepackage{natbib}

\slugcomment{Accepted for publication on ApJ}

\shorttitle{Millimeter imaging of MWC~758}
\shortauthors{Isella A., Natta A., Wilner D., Carpenter J.M., and Testi L.}

\begin{document}


\title{Millimeter imaging of MWC~758: probing the disk structure and kinematics}
 

\author{Andrea Isella}
\affil{Department of Astronomy, California Institute of Technology, MC 249-17, Pasadena, CA 91125, USA}
\email{isella@astro.caltech.edu}
\author{Antonella Natta}
\affil{Osservatorio Astrofisico di Arcetri, INAF, Largo E. Fermi 5, 50125 Firenze, Italy} 
\author{David Wilner}
\affil{Harvard-Smithsonian Center for Astrophysics, 60 Garden Street, MS 42, Cambridge, MA 02138, USA}
\author{John M. Carpenter}
\affil{Department of Astronomy, California Institute of Technology, MC 249-17, Pasadena, CA 91125, USA}
\author{Leonardo Testi}
\affil{European Southern Observatory, Karl Schwarzschild str. 2, D-85748 Garching, Germany}


\begin{abstract}
We investigate the structure and kinematics of the circumstellar disk around the Herbig Ae star MWC~758 
using high resolution observations of the $^{12}$CO (3-2) and dust continuum emission at 
the wavelengths of 0.87 and 3.3~mm. We find that the dust emission peaks at an orbital 
radius of about 100 AU, while the CO intensity has a central peak coincident with the position of the star.
The CO emission is in agreement with a disk in keplerian rotation around a 2.0~M$_\sun$ star, confirming 
that MWC~758 is indeed an intermediate mass star. By comparing the observation with theoretical disk 
models, we derive that the disk surface density $\Sigma(r)$ steeply increases from 40 to 100 AU, 
and decreases exponentially outward. Within 40 AU, the disk has to be optically 
thin in the continuum emission at millimeter wavelengths to explain the observed dust morphology, though our observations 
lack the angular resolution and sensitivity required to constrain the surface density on these spatial scales.
The surface density distribution in MWC~758 disk is similar to that of ``transition'' disks, 
though no disk clearing has been previously inferred from 
the analysis of the spectral energy distribution (SED). Moreover, the asymmetries observed in the dust and 
CO emission suggest that the disk may be gravitationally perturbed by a low mass companion orbiting
within a radius of 30 AU.
Our results emphasize that SEDs alone do not provide 
a complete picture of disk structure and that high resolution millimeter-wave images are essential to reveal 
the structure of the cool disk mid plane.

\end{abstract}



\section{Introduction}

Millimeter-wave observations of circumstellar disks provide the best tool to measure 
the distribution and kinematics of the circumstellar material.  These observations have 
for years achieved 1-3\arcsec\ angular resolution to resolve the disk emission on 
scales down to an orbital radius of about $\sim$70 AU at the distance of the nearby 
star forming regions \citep[e.g.][]{Koerner95,Guilloteau98,Kitamura02}. More recent 
observations are reaching subarcsecond resolution that are resolving disks on scales 
of the Kuiper Belt and yielding unprecedented informations on the innermost disk regions 
\citep[see, e.g,][]{Andrews09,Brown09,Hughes09,Isella09,Isella10,Pietu06}.

In many cases millimeter observations confirm the radial distribution of dust and gas 
inferred from the analysis of the disk spectral energy distribution (hereafter SED).
The large majority of pre-main sequence stars exhibit a flux excess over the 
stellar photosphere from near-infrared to centimeter wavelengths. This excess arises 
from a ``classical'', or primordial, disk that extends from few stellar radii to few hundreds 
of astronomical units. However, about 10\% of the observed systems \citep{Muzerolle10} 
exhibit a deficit of flux in the near- and mid-infrared compared to a ``classical'' disk, 
but have similar level of far-infrared emission \citep[i.e., ``transition'' disk;][]{Strom89}. 
The SED shape suggests that these stars lack the emission from warm dust close to the central 
star. The lack of warm dust emission may result from several different effects such as the 
presence of a stellar companion, the formation of a planetary system, 
disk photoevaporation, magneto-rotational instabilities, 
disk viscosity, and grain growth \citep{Alexander07,Chiang07,Ireland08,Isella09,Strom89,Tanaka05}. 
Subsequent high-resolution millimeter-wave images have confirmed that several ``transition'' 
disks are indeed dust depleted within the radius predicted by the SED modeling \citep{Pietu06,Brown09,Hughes09}.

The increasing sensitivity and resolution of (sub)millimeter-wave interferometers 
is now providing unexpected insights about the structure of ``classical'' disks.
 Sub-arcsecond observations of AB~Aur  and RY~Tau systems show indeed that these two
circumstellar disks may be dust depleted within orbital radii of 70 AU and 13 AU 
respectively \citep{Pietu05,Isella10}, although no disk clearing has been inferred from the 
analysis of the SED \citep{Robitaille07,Dullemond01}. It is therefore possible that other  
``classical'' disks may be characterized by a disk surface density that drastically deviates 
from that inferred from the SED modeling. In particular, since the dust opacity in the infrared
is much larger than at millimeter-wavelengths, it is reasonable to assume that a large fraction of ``classical" 
disks might by characterized by partially cleared inner cavities which are not detectable 
through the analysis of the infrared SED.

In this paper we discuss the Herbig Ae star MWC 758 (HD 36112),  a 3~Myr old 
pre-main sequence A5 star located near the edge of the Taurus star forming region at a distance 
of 200$^{+60}_{-40}$ pc \citep{Perryman97,vandenAncker98}. Previous investigations based on 
the SED modeling and on spatially resolved interferometric observations suggested that MWC~758 
is surrounded by a classical, fully flared rotating disk that extends from few stellar radii \citep{Isella08,Eisner04} 
up to an outer radius of $385\pm26$ AU\footnote{Note that Chapillon et al. and Eisner et al. assume a
stellar distance of 140 and 150 pc respectively. The radii quoted from these two papers have been therefore
rescaled for the distance of 200 pc} \citep[][]{Chapillon08}. At 1.3~mm, dust continuum emission 
observed at 2\arcsec\ angular resolution is centrally peak and more compact than the $^{12}$CO (2-1) emission 
\citep[][]{Chapillon08}. This difference in radial extension between dust and gas emission is similar to 
that found in other circumstellar disks \citep[see, e.g.,][]{Pietu05,Isella07} and can be explained with 
an exponential decrease of the disk surface density in the outermost disk regions \citep[][]{Hughes08, Isella10}.  

Here we present new  images of the millimeter-wave dust continuum 
and $^{12}$CO emission toward MWC~758 characterized by an angular resolution of 0.7\arcsec, which is 
three times better than previous observations.  Our observations resolve the inner disk into 
two peaks in the continuum emission, strongly suggesting that the disk surface density 
decreases within an orbital radius of about 100 AU. Similarly to the cases of AB~Aur and RY~Tau, the MWC~758 inner 
disk may be dust depleted with respect to a power law surface density profile, though no 
disk clearing has been inferred from the analysis of the SED.

The observations, presented in Section~\ref{sec:obs} and ~\ref{sec:res}, are compared with theoretical disk 
models to constrain the surface density profile of MWC~758 disk. The results are presented in 
Section~\ref{sec:analysis}. The discussion and conclusions follow in Section~\ref{sec:disc}.

 \section{Observations and data reduction}
\label{sec:obs}
 
 %
  

\subsection{SMA observations}

\begin{figure}[!t]
\centering
\includegraphics[angle=0,width=\columnwidth]{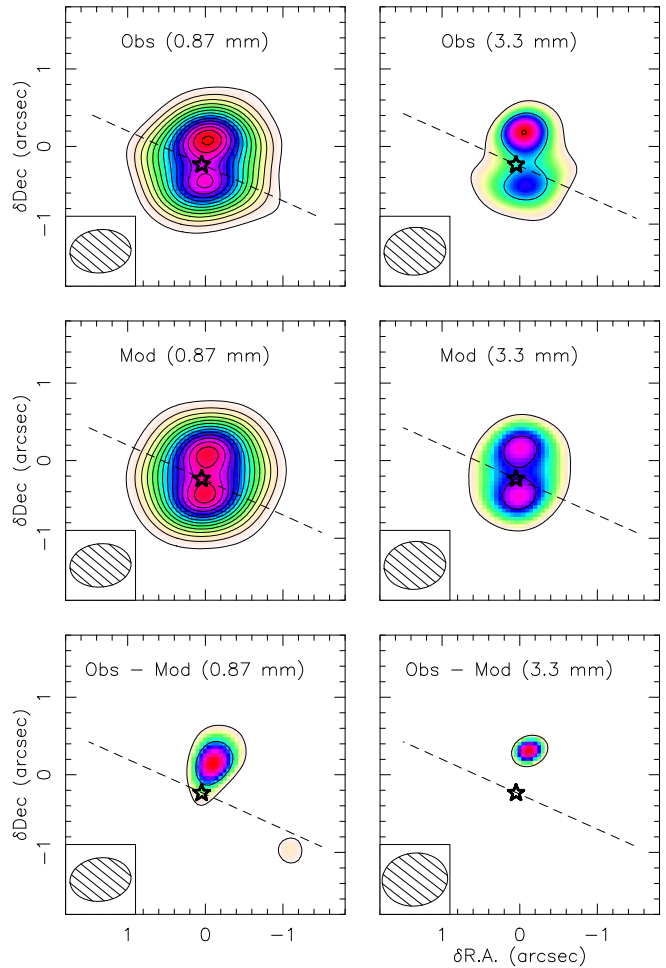}
\caption{\label{fig:cont} Observations, best fitting models and residual maps of the dust 
continuum emission observed toward MWC~758 by SMA at 0.87 mm (left column) 
and by CARMA at 3.3 mm (right colums). The 1$\sigma$ noise levels are 2.1 
and 0.25 mJy/beam in SMA and CARMA observations respectively. Contours
are plotted every $3\sigma$. The symbol $\star$ marks the position of MWC~758 after
correcting for the proper motion as described in Section~\ref{sec:res}. The dashed
line shows the orientation of the disk major axis derived from the CO observations 
as discussed in Section~\ref{sec:res}.
}
\end{figure}

\begin{figure*}[!t]
\centering
\includegraphics[angle=270,width=1.0\textwidth]{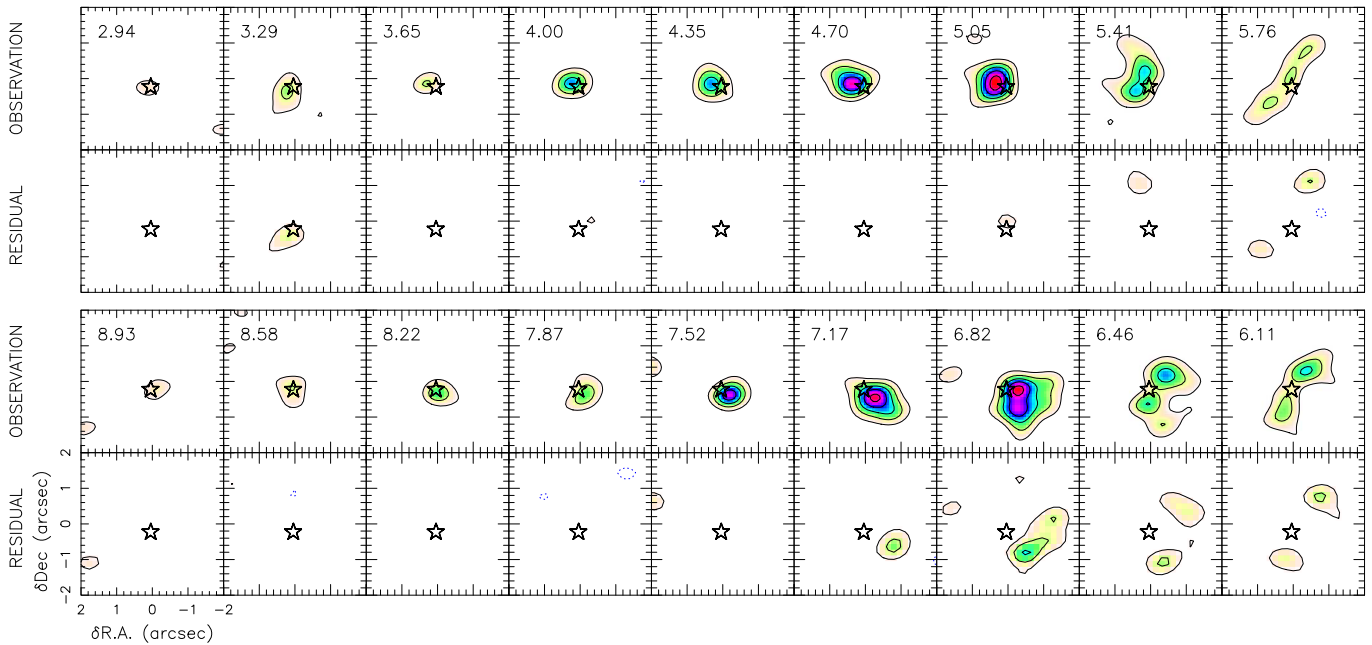}
\caption{\label{fig:CO} Channel maps for the $^{12}$CO J=3-2 line emission observed toward MWC 758 in the velocity range between 2.94 and 8.93 km/sec. The contours are spaced by $3\sigma=0.25$ Jy/beam. The figure also shows the residuals obtained by subtracting the best fitting model CO emission from the observations. The symbol $\star$ marks 
the position of the central star after correcting for the proper motion.
}
\end{figure*}

The MWC~758 system was observed at 345 GHz with the Submillimeter 
Array\footnote{The Submillimeter Array is a joint project between the Smithsonian 
Astrophysical Observatory and the Academia Sinica Institute of Astronomy and 
Astrophysics, and is funded by the Smithsonian Institution and the Academia Sinica.} (SMA)
on Mauna Kea, Hawaii, on 2008 Jan 28. The array was in the extended
configuration, and the seven available antennas provided baseline lengths
from 31 to 225 meters. The atmospheric transparency was excellent through
the track, with an opacity of 0.04 at 225~GHz measured at the nearby Caltech
Submillimeter Observatory.
The correlator provided 2~GHz of bandwidth in two sidebands separated by
10~GHz, with the upper sideband centered near the CO J=3-2 line at 345.796~GHz.
This line was observed at a spectral resolution 0.18~km~s$^{-1}$ in a single
correlator ``chunk'' spanning 104~MHz. The primary flux calibrator was Titan,
and the passband calibrators were the quasars 3C273 and 3C111.
Complex gain calibration was carried out with observations of the quasar
J0530+135 (1.75~Jy) interleaved with MWC~758. Data was obtained for MWC~758
over the hour angle range $-3.2$ to $+4.7$. All of the data editing and
calibration were performed using standard routines in the MIR software package
for IDL. The resulting map of the dust continuum emission is shown in Figure~\ref{fig:cont}. Applying 
natural weighting to the data we achieve a noise level of 2.1~mJy/beam and 
synthesized beam FWHM of 0.76\arcsec$\times$0.56\arcsec. Figure~\ref{fig:CO}
shows the CO emission in the velocity range between 2.94 and 8.93 km s$^{-1}$, 
as obtained by averaging the observations every two channels. The resulting noise 
level is 83~mJy/beam.

\subsection{CARMA observations}

The dust thermal emission toward MWC~758 was observed on January 2009 
using the Combined Array for Research in Millimeter-wave Astronomy\footnote{Support 
for CARMA construction was derived from the Gordon and Betty Moore Foundation, the 
Kenneth T. and Eileen L. Norris Foundation, the James S. McDonnell Foundation, the 
Associates of the California Institute of Technology, the University of Chicago, the states 
of California, Illinois, and Maryland, and the National Science Foundation. Ongoing 
CARMA development and operations are supported by the National Science Foundation 
under a cooperative agreement, and by the CARMA partner universities.} (CARMA)
located at the altitude of 7200 feet in the Inyo mountains of eastern California. The 
array was in the extended B configuration, which provides baselines between 80.9 and 
946.3~m.  The observations were obtained at the local oscillator frequency of 
97.5~GHz ($\lambda \sim 3.3$~mm). The CARMA correlator was configured with six 
bands, each of which was configured to 468 MHz band-width to provide 
maximum continuum sensitivity. No molecular emission line was observed 
with CARMA during this track. The passband shape was calibrated by observing  3C454.3 and the 
flux calibration was set by observing Uranus. The quasar J0510+180 was observed 
every 15 minutes to correct for atmospheric and instrumental effects. The data 
reduction and the image reconstruction were obtained using the MIRIAD software 
package. Figure~\ref{fig:cont} shows the resulting map, which is 
characterized by a rms noise of 0.25~mJy/beam and a synthesized beam FWHM 
of 0.80\arcsec$\times$0.62\arcsec\  when natural weighting is applied to the data.

\section{Results}
\label{sec:res}

\begin{figure*}[t]
\centering
\includegraphics[angle=270,width=\textwidth]{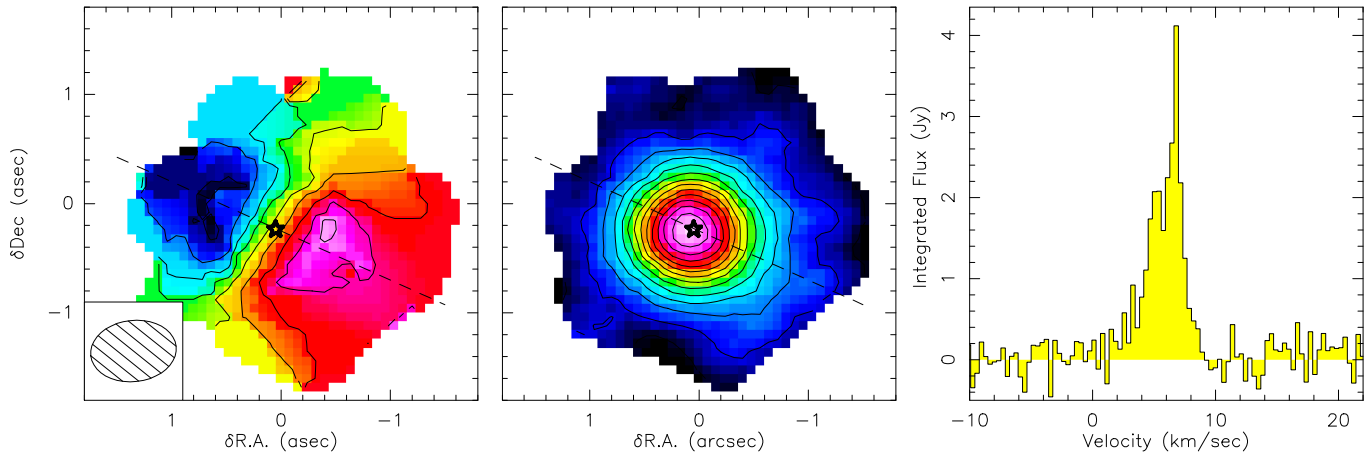}
\caption{\label{fig:CO_mom} 
{\it Left panel}: $^{12}$CO J=3-2 velocity gradient across MWC~758 circumstellar disk. Red 
and blue colors indicate the red- and blue-shifted disk velocities respectively. Velocity contours 
start at 4.7 km sec$^{-1}$ and are spaced by 0.3 km sec$^{-1}$.  The dashed line indicates 
the position angle of the apparent disk major axis, which is rotated by 65\arcdeg\ to east 
from north. {\it Central panel} $^{12}$CO J=3-2 integrated intensity map.
{\it Right panel:} $^{12}$CO J=3-2 spectrum obtained integrating over the region shown in the 
intensity map. 
}
\end{figure*}

Figures~\ref{fig:cont}, \ref{fig:CO}, and \ref{fig:CO_mom} show the dust and 
CO emission towards MWC~758 as observed by pointing the telescope at 
the ICRS coordinates of the star at the epoch J2000, RA=05h 30m 27s.530 ($\pm12.89$ mas) and  
Dec=+25\arcdeg\ 19\arcmin\ 57\arcsec.08 ($\pm5.57$ mas).
The position of the star at the date of the observation is shown by the symbol $\star$  
and is calculated by applying a proper motion of  $\delta\textrm{(R.A)}$=5.23$\pm$1.40 mas/yr and 
$\delta\textrm{(Dec)}$=-25.95$\pm$0.66 mas/yr to the J2000 coordinates \citep[both the stellar coordinates and 
the proper motion are from][]{Perryman97}.
The uncertainties on the stellar position resulting both from the Hypparcos astrometry and 
proper motion uncertainties are smaller than the size of symbol itself.

The dust emission at both 0.87 and 3.3 mm is characterized by two peaks which are 
symmetric with respect of the position of the star and are separated by about 0.7\arcsec, 
or 140 AU at the distance of 200~pc.
At the wavelength of 0.87 mm, the northern and southern peaks have fluxes 
of 89 and 77 mJy/beam respectively. A similar flux ratio between 
the two peaks is measured at 3.3~mm, suggesting that the observed morphology 
corresponds to an asymmetric dust emission between the two sides of the disk. 
By integrating over a circle with a radius of 1.5\arcsec\ centered at the position 
of the star, we measure fluxes of 217$\pm$40 mJy and 4.9$\pm$1.7 mJy at 
0.87 and 3.3~mm respectively, where the errors account for both the thermal noise 
and absolute calibration uncertainties. The integrated fluxes are in agreement with the
values measured at 2\arcsec\ angular resolution by \citet{Chapillon08} suggesting that 
the dust emission is all coming from the spatial scales probed by our observations. 
Assuming that the disk millimeter emission follow a power law $F(\nu)\propto \nu^\alpha$, we derive 
a spectral index $\alpha_{\textrm{1-3mm}}$=3.0. 

Figure~\ref{fig:CO} shows the \coline\ channel maps in the velocity range between 2.94
and 8.93~km/sec. The CO emission is spatially resolved and shows a velocity pattern 
typical of an inclined rotating disk. The first moment of the CO emission, i.e., the velocity 
gradients across the disk, indicates that the western and eastern sides of the disk 
are respectively red- and blue-shifted with respect to the relative velocity of the 
star (see the left panel of Figure~\ref{fig:CO_mom}). The velocity gradient also 
suggests a position angle of the disk major axis of about 65\arcdeg\ as 
measured from north to east.
The central panel of Figure~\ref{fig:CO_mom} shows the spectrally integrated CO emission.
In contrast to what is observed in the dust continuum, the CO emission has a single peak coincident with the 
position of the star.  The CO emission is observed up to a distance of 1.3\arcsec\ in the red-shifted
side of the disk, and only to about 1\arcsec\ in the other side. 
The line integrated flux is 12.9 Jy km/sec. This value is about 
14\% smaller than that measured in single dish observations \citep{Dent05}, suggesting that 
part of the CO emission is resolved out in the interferometric observations. As a consequence,  the 
observed asymmetry in the CO emission might be due to the lack of short baselines 
in the SMA observations instead of to a real asymmetry in the gas temperature or distribution.
Finally, the spatially integrated CO spectrum (right panel of 
Figure~\ref{fig:CO_mom}) has a double peak profile centered at the velocity of $v_{lsr}=6.0$ km/sec. 
The line emission is observed in the velocity range $v_{lsr}\pm$4 km/sec with a brighter flux in 
the red-shifted half of the disk. 

To date, the different morphologies observed in the dust and gas emission toward 
MWC~758 represent a unique case among the sample of circumstellar disks 
spatially resolved with millimeter-wave interferometers. Our observations suggest that 
the optical depth of the dust continuum emission decreases within a radius of 
70-100 AU from the star. Despite that,  the disk remains gas-rich down to at least a 
distance of 30~AU to explain the centrally peaked CO emission. In addition, 
the SED of MWC~758 shows a flux excess above the stellar photosphere 
between near- and far-infrared wavelengths typical of optically thick flaring 
disk extending to few stellar radii (see Figure~\ref{fig:sed}). The next section is devoted 
to reconciling all these observations in the framework of existing disk models, in order
to constrain the radial distribution of the circumstellar material around MWC~758.


\section{Data analysis}
\label{sec:analysis}

\subsection{Disk models}
\label{sec:model}
To analyze the data we adopt the disk 
model discussed in \citet[][]{Isella07,Isella09}. In brief, we compute the 
radial dust temperature profile following the "two-layer" approximation of \citet{Chiang97}, where 
the disk is characterized by a warm surface layer and a cooler interior. Both temperatures are calculated 
as a function of the orbital radius by iterating on the vertical disk structure \citep[see][]{Dullemond01}. 
The disk is in hydrostatic equilibrium between the gas pressure and the stellar gravity, which leads to 
a flared geometry with the opening angle increasing with the distance from the central star.

The dust opacity is calculated by assuming an interstellar grain composition \citep{Pollack94} and a 
particle size distribution between a minimum and a maximum value $a_{min}$ and $a_{max}$ 
according to $n(a) \propto a^{-q}$. We fix $a_{min} = 0.05$~$\mu$m and vary $a_{max}$ and $q$ 
to reproduce the spectral index $\alpha$ of the millimeter disk emission. For sake of simplicity 
we assume that the dust opacity is constant throughout the disk \citep[although see][]{Birnstiel10b}.

The radial distribution of the circumstellar material follows the similarity solution for the disk surface 
density of a viscous keplerian disk \citep{Lynden74} expressed by 
\begin{equation}
\label{eq:sigma_used}
\Sigma(r,t) = \Sigma_t \left( \frac{r_t}{r} \right)^{\gamma} \times  
\exp{ \left\{ - \frac{1}{2(2-\gamma)} 
\left[ \left( \frac{r}{r_t} \right)^{(2-\gamma)} -1 \right] \right\} },
\end{equation}
where the characteristic radius $r_t$, $\gamma$, and the surface density normalization 
$\Sigma_t$ are free parameters of the model, as well as the disk inclination and position angle.
  
From the derived dust density, temperature, and opacity we calculate the disk SED and synthetic disk 
images in the dust continuum at 0.87~mm and 3.3~mm using the radiative transfer 
solution discussed in \citet{Dullemond01}. The synthetic disk images are then Fourier transformed and 
sampled at the appropriate positions on the (u,v) plane corresponding to our CARMA and SMA observations. 

The CO emission is calculated assuming that rotational levels are thermalized. 
The CO temperature is parameterized as a function of the radius as $T_{\textrm{CO}}(r) 
= T_{\textrm{CO}}(r_0)(r_0/r)^\zeta$, where $T_{\textrm{CO}}(r_0)$ and $\zeta$ are free 
parameters and $r_0=50$ AU. The CO density is then calculated by assuming that
the dust and the gas are well mixed with a gas-to-dust ratio of 100. The adopted CO 
abundance is $^{12}$CO/H$_2 = 7.0 \times 10^{-5}$. Finally, the CO channel maps are
calculated in the velocity range probed by our observations by assuming that the emission 
originates from an inclined keplerian disk. Both the disk inclination and the mass of the 
central star are free parameters of the model \citep[see the Appendix of ][ for more 
details on the CO emission model]{Isella07}. As for the case of the dust emission, 
the synthetic CO channel maps are then Fourier transformed and compared to the 
measured complex visibilties.

The best fitting models for both the dust and gas emission, and the uncertainties on the model 
free parameters, are then obtained by $\chi^2$ minimization carried out using  the Markov Chain Monte Carlo 
method discussed in \citet{Isella09}.


\subsection{Results of the model fitting}
\label{sec:res_mod}


\begin{table}
\caption{\label{tab:model} Best fitting model parameters for the dust and CO millimeter emission observed towards MWC~758.  The
quoted  uncertainties correspond to the 1$\sigma$ uncertainty level \label{tab:dust}}
\centering
\begin{tabular}{l}

\hline
Disk orientation:  \\
Inclination = 21\arcdeg$\pm$2\arcdeg, P.A. =  65\arcdeg$\pm$7\arcdeg  \\
\hline
Surface density $\Sigma(R)$:\\  
$\gamma$ = -4.1$\pm0.3$, $R_t$ = 69$\pm$3 AU,  $\Sigma_t$ =1.5$\pm$0.1 g cm$^{-2}$ \\ 
\hline
CO temperature $T(\textrm{CO})$:\\
 T$_{\textrm{CO}}$(50 AU)  = 53$\pm$4 K, $\zeta$ = -0.1$\pm$0.2 \\
\hline
M$_\star$ = 2.0$\pm$0.2 M$_\sun$  \\
\hline
Grain size distribution $n(a)$:\\
 $a_{max}$ = 1 cm, q = 3.5  \\
\hline
\end{tabular}
\end{table}

\begin{figure}
 \centering
 \includegraphics[angle=0,width=\columnwidth]{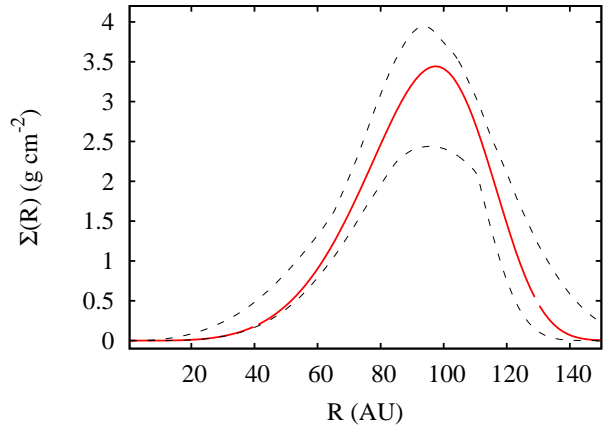}
 \caption{\label{fig:sigma} The solid red line shows the best fitting model solution for the disk surface 
 density $\Sigma(r)$ obtained by analyzing the spatially resolved observations of the dust
 and CO emission at both 0.87 and 3.3~mm. The dashed lines indicate the 3$\sigma$ 
 uncertainty range. Given the angular resolution and sensitivity 
 of our observations, $\Sigma(r)$ is constrained only between 40 and 130 AU.
 Within 40 AU, $\Sigma(r)$ can significantly deviate from the profile shown in the figure, 
 though the disk must remain optically thin at both 0.87 and 3.3~mm to explain the two
 peaks observed in the dust emission.
}
\end{figure}

A satisfactory fit of the spatially resolved dust and CO emission, as well as
of the SED between far-infrared and (sub)millimeter wavelengths is obtained with the disk model 
presented in Table~\ref{tab:model}. The disk surface density is mainly constrained 
by the optically thin dust emission at 0.87 and 3.3~mm and is characterized by $\gamma=-4.1$ 
and $r_t=69$ AU. As shown in Figure~\ref{fig:sigma}, $\Sigma(r)$ has a "bell" shape 
that reaches the maximum value of 3.3 g/cm$^2$ at the radius of 100 AU. For $r \ll r_t$,  
the surface density is proportional to $r^4$, suggesting that a steep radial density profile is 
required to explain the two peaks observed in the dust emission.  For $r>100$ AU, $\Sigma(r)$
fades exponentially as prescribed by Equation~\ref{eq:sigma_used}. Given the angular resolution 
and sensitivity of our observations, we can constrain the surface density only between 40 
and 130 AU. In this range, the disk is optically thin to the dust continuum emission at both 
0.87 and 3.3~mm. Within 40 AU, the radial profile of $\Sigma(r)$ might significantly deviate 
from that shown in  Figure~\ref{fig:sigma}. However, the disk must remain optically thin to the 
dust continuum millimeter-wave emission down to at least few AU in order to reproduce 
the two peaks shown in Figure~\ref{fig:cont}. 

Radially integrating the surface density we obtain a disk mass of 0.01 M$_\sun$. This 
result is affected  by a large error due to the uncertainties on the dust opacity and on the gas to dust ratio. 
In particular, the analysis of the $^{12}$CO (2-1) and (1-0) line emission suggests that 
the gas to dust ratio in MWC~758 disk might me smaller than 100 \citep{Chapillon08}, 
leading therefore to a smaller disk mass.
 
Despite the steep surface density profile, the $^{12}$CO (3-2) 
emission remains optically thick down to a radius of about 20~AU. 
As a consequence, the CO emission does 
not constrain the disk surface density but it provides a measure
of the CO temperature in the form $T_{\textrm{CO}}(r) \simeq 53\textrm{K} \times (50 \textrm{AU}/r)^{-0.1}$.
Note that the $^{12}$CO (3-2) emission remains optically thick even if the CO is largely depleted with respect
to the standard abundance. In particular, assuming a CO abundance of $10^{-6}$ \citep{Chapillon08} we calculate 
a mean optical depth of about $10^3$. 
The almost radially constant profile of the CO temperature is in good agreement with
the value $\zeta=0.05\pm0.2$ derived from the analysis of the $^{12}$CO (2-1) emission line 
by \citet{Chapillon08}. In the assumption that the disk is in keplerian rotation, we 
derive a disk inclination of 22\arcdeg\ and a dynamical mass for the central star of 
2.0~M$_\sun$, similar to the value of 2.3~$M_\sun$ derived from the 
H-R diagram using stellar evolutionary tracks \citep{Manoj06}.

The residuals obtained by subtracting the best fitting model to the observations reveal 
asymmetries in the MWC~758 disk. In the case of the dust emission 
(see Figure~\ref{fig:cont}), our disk model fails to reproduce the flux measured 
in the northern side of the disk at both 0.87 and 3.3~mm. From the intensity of the residuals 
we calculate that the dust density, opacity,  and/or temperature in the northern side of 
the disk might be 20\% higher than the southern side.
Figure~\ref{fig:CO} shows the residuals for the CO channel maps in the velocity range between 
2.94 and 6.11 km s$^{-1}$. Most of the significative structures appear in low-velocity channels of 
the red-shifted side of the disk between 6.11 and 7.11 km/sec. Since the CO emission is 
almost completely optically thick, this suggest that the CO temperature might be higher in 
this side, or that the CO extends to slightly larger distances from the central star.
Note, however, that the CO asymmetry might also be due to the lack of short spatial
frequency in SMA observations as discussed in Section~\ref{sec:res}.

\begin{figure}
 \centering
 \includegraphics[angle=0,width=\columnwidth]{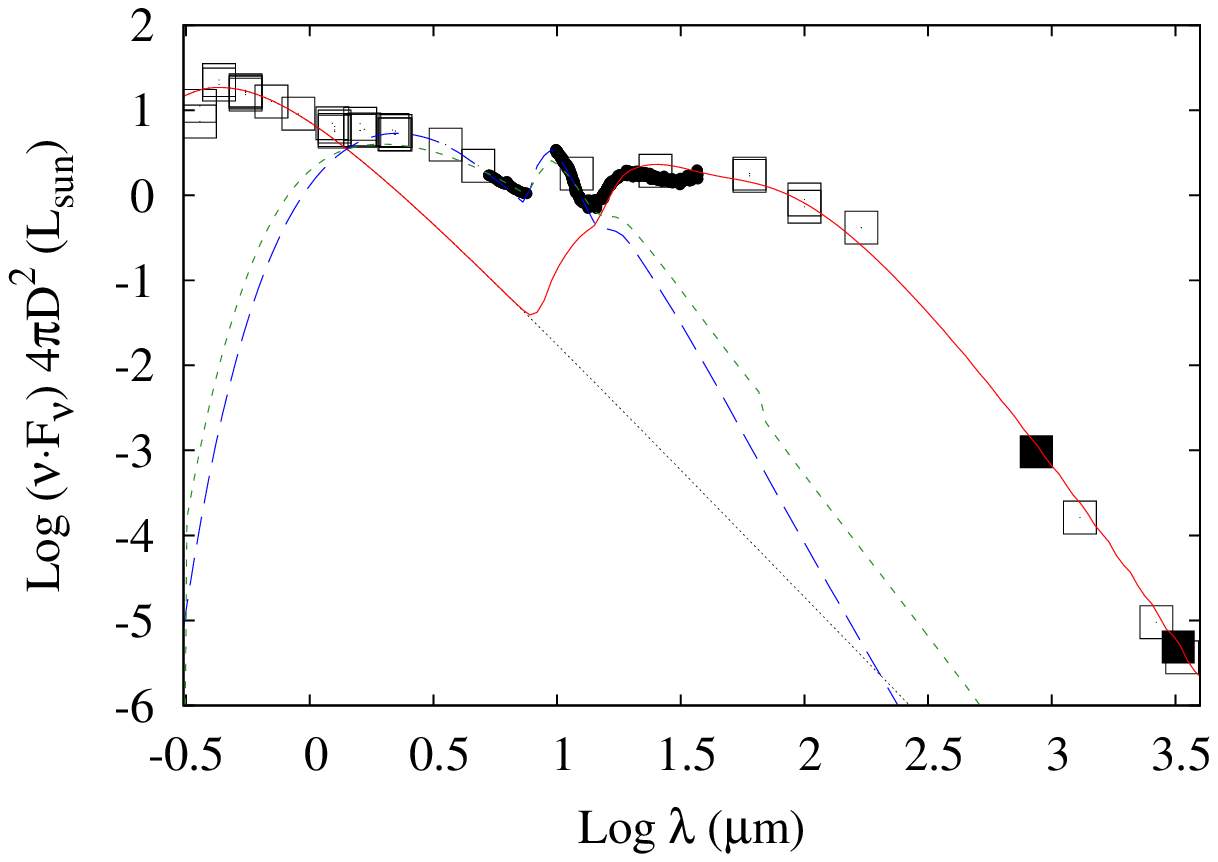}
  \caption{\label{fig:sed} Spectral energy distribution of MWC~758. Photometric data from the literature 
  are shown with empty squares \citep{Eisner04,Malfait98,Beskrovnaya99,Cutri03,IRAS,Elia05,Chapillon08}.
  The thick black line show the observed IRS spectrum from the Spitzer archive. The filled squares show 
  our new measurements at 0.8 
  and 3.3~mm. Optical and near-infrared data were de-reddened adopting 
  $A_V=0.2$~\citep{vandenAncker98}.  The dotted black line shows the stellar photosphere 
  modeled using a black body emission with a temperature $T_{eff}=8200$~K, 
  $L_\star=22$ L$_\sun$, and a distance of 200~pc \citep{vandenAncker98}. 
  The solid red line shows the SED for the best fitting model discussed in Section~\ref{sec:res}. 
  The long-dashed blue and short-dashed green curves show the SED of an optically thick and thin 
  inner disk model, as discussed in Sec.~\ref{sec:res_mod}.}
 \end{figure}

Figure~\ref{fig:sed} shows the comparison between the observed SED and the SED of the best 
fitting model to the millimeter data. At millimeter wavelengths, a good agreement with the observations 
is obtained with a grain size distribution slope $q=3.5$ and a maximum grain size $a_{max} = 1$ cm. 
Fitting the dust opacity $k(\lambda)$ between 1.3 and 2.8~mm with a power law $k(\lambda) \propto \lambda^{-\beta}$, 
we obtain $\beta=1.35$, which is a factor of two higher than the mean value measured in T Tauri stars but 
in agreement with those found in other Herbig Ae stars \citep[see, e.g.,][]{Ricci10,Natta04}. 

  The good agreement between the model and the 
observation extends down to a wavelengths of about 20~$\mu$m. This suggests that the disk has
a flared structure in the region that emits most of the mid- and far-infrared emission, namely
between 20 and 110 AU.  At shorter wavelengths, the emission of the best 
fitting model is  much lower than what observed toward MWC~758. This is due to the fact that 
for $\Sigma(r) \propto r^4$,  the dust density within few AUs from the central star is 
lower than $10^{-6}$ g cm$^{-2}$ and its near-infrared emission is negligible with respect to 
the stellar photospheric emission. The observed near-infrared excess therefore suggests the 
presence of material with temperature between 1000 and 2000 K in excess to what predicted 
by our model. 

To reconcile the near-infrared SED with our millimeter-wave observations we propose
two different scenarios. In the first case, the near-infrared emission is coming from
an optically thick disk extending between 0.05 and 5 AU. This model comprises 
a gaseous disk between 0.05 and 0.1~AU, a "puffed-up" inner rim of the dusty disk between 0.4 and
 0.5 AU \citep[see, e.g,][]{Isella05}, and a flaring disk between 0.5 and 5 AU. As discussed in \citet{Isella08}, 
 this model explains the spectro-interferometric observations of the H- (1.6~$\mu$m) and 
 K-band (2.2~$\mu$m) disk emission \citep[see also][]{Benisty10}. The resulting SED, which 
 is indicated by the short-dashed green line in Figure~\ref{fig:sed}, provides a good 
 description of the near-infrared emission and of the 10~$\mu$m silicate feature.  
Alternatively, the infrared SED can be explained with the presence of an optically thin 
layer of dust surrounding the star \citep{Vinkovic06, Mulders10}. In this case, we calculate 
that a ring of dust extending between 0.15 and 0.8 AU, and characterized by an optical 
depth for the stellar radiation along the direction perpendicular to the disk mid plane 
of 0.4 is required to reproduce the disk emission (see the long-dashed curve in 
Figure~\ref{fig:sed}). For the assumed dust opacity, this model requires a minimum surface 
density of 0.01 g cm$^2$, or a total mass of $2\times10^{-9}$ M$_\sun$. 
Note that neither of these models will produce a significant dust and CO emission
at millimeter wavelengths. 

To summarize, assuming the similarity solution for a keplerian viscous disk, 
we find strong evidence that the surface density of the MWC~758 disk
increases  as $r^4$ between 40 and 100 AU from the central star, and 
decreases exponentially outwards. However, this solution for the surface density is not
unique and observations with higher angular resolution are required to discriminate between 
different models. We note, for example, that a discontinuity in the surface density at about 80 AU 
from the central star might provide an equally good fit with a different step gradient. 
Within $40$ AU,  $\Sigma(r)$ is constrained by the analysis of 
the SED.  We calculate that a minimum surface density value of 10$^{-2}$ g cm$^{-2}$ between 0.15 
and 0.8 AU is required to reproduce the observed near- and mid-infrared emission. 
Probing these spatial scales  requires high sensitivity 0.1\arcsec\ resolution 
observations that will become possible with the Atacama Large Millimeter Array (ALMA).

\section{Discussion and conclusions}
\label{sec:disc}
The surface density profile in MWC~758 is similar to that measured in the 
transitional disk around the T Tauri star LkCa~15.  In this case $\Sigma(r)$
peaks at about 50 AU from the central star \citep{Pietu06,Espaillat07} and a minimum 
mass of $1\times10^{-11}$ M$_\sun$ within 5 AU from the central star is required to 
produce the observed near-infrared emission \citep{Mulders10}. Similarly to what suggested 
for LkCa15, the shape of the surface density in MWC~758 may be due to the disk clearing 
operated by a low mass stellar companion, or by one or more giant planets. To this regard, we note 
that recent spectroastrometric observations of MWC~758 exclude the presence of  a stellar companion 
more massive than about 0.9 M$_\sun$ orbiting between 0.15\arcsec\ and 0.3\arcsec, or  30 and 60 AU for 
the assumed stellar distance of 200 pc \citep[][Wheelwright H. 
private communication]{Wheelwright10}. The nature of a possible companion of MWC~758 remains 
therefore uncertain. 

The asymmetries observed in the dust and CO emission support the hypothesis that the disk may 
be gravitationally perturbed by a massive body. The asymmetric CO emission can be due to a warped 
optically thick inner disk, i.e., more inclined than the 
outer disk, so that the CO in north-west side of the disk is partially shielded by the inner 
disk from the stellar radiation. This would result in a lower gas temperature and consequently 
lower emission with respect to the other side of the disk \citep[see, e.g.,][]{Panic10}. Indeed,
we note that  the inclination of the inner disk derived from near-infrared interferometric 
observations  is of about 40\arcdeg$\pm$10\arcdeg\ \citep{Isella06,Isella08,Eisner04}, which 
is larger than the 21\arcdeg$\pm$2\arcdeg\ derived in this work. 

Other mechanisms, which are in principle able to evacuate the inner disk, do not 
seem to be consistent with the observations. Photoevaporation and magneto rotational
instability models predict completely cleared inner disks and  surface density 
profiles at the inner disk radius much steeper than $r^4$ \citep{Alexander07,Chiang07}.
In addition, the surface density within few AU is much larger than that predicted by the viscous 
solution for the surface density that best fits the millimeter data. This fact excludes that $\Sigma(r)$ is 
shaped only by the radial profile of the disk viscosity \citep[see the discussion in ][]{Isella09}.  
Low dust opacity caused by grain growth has also been suggested to explain inner 
holes observed in the continuum emission at millimeter wavelengths \citep{Tanaka05}. 
In the case of MWC~758, the dust opacity should decrease by at least two 
orders of magnitude to explain the observations.  This would require a grain size distribution 
flatter than what typically assumed (i.e., $n(a) \propto a^{-q}$ with $q < 3$) 
and a steep radial decrease of the maximum grain (i.e, form 10 cm at 10 AU to 10~$\mu$m at 100 AU).

In conclusion, our observations show that the disk around MWC~758 is characterized 
by a steeply increasing surface density  between 40 and 100 AU from the central star.
Given the infrared SED and the asymmetries observed in the gas and dust emission, we
conclude that the surface density might be shaped by the presence of stellar, or planetary 
size companion orbiting within 30 AU. The fact that MWC~758 was previously labeled 
as ``classical'' disk emphasizes 
that SEDs alone do not provide a complete picture of disk structure and that high resolution
millimeter-wave images are essential to reveal the structure of the cool disk mid plane.

\acknowledgments
We thank the OVRO/CARMA staff and the CARMA observers for their assistance 
in obtaining the data. We acknowledge support from the Owens Valley Radio Observatory, 
which is supported by the National Science Foundation through grant AST 05-40399.
This work was performed in part under contract with the Jet Propulsion Laboratory 
(JPL) funded by NASA through the Michelson Fellowship Program. JPL is 
managed for NASA by the California Institute of Technology.


\bibliographystyle{apj}
\bibliography{ref}

	
\end{document}